\documentclass[fleqn,usenatbib]{mnras}

\usepackage[T1]{fontenc}
\usepackage{ae,aecompl}

\usepackage{longtable}

\usepackage{graphicx}	
\usepackage{amsmath}	
\usepackage{amssymb}	

\title[G4.8+6.2, kilonova remnant]{G4.8+6.2, a possible kilonova remnant?}

\author[Y. Liu et al.]{
Yu Liu,$^{1,3}$
Yuan-Chuan Zou,$^{1}$\thanks{E-mail: zouyc@hust.edu.cn}
Bing Jiang,$^{2}$
He Gao,$^{3}$
Shuai-Bing Ma,$^{1}$
Bin Liao$^{1}$
\\
$^{1}$School of Physics, Huazhong University of Science and Technology, Wuhan 430074, China (yul@hust.edu.cn, zouyc@hust.edu.cn)\\
$^{2}$School of Astronomy and Space Science, Nanjing University, Nanjing 210093, China \\
$^{3}$Department of Astronomy, Beijing Normal University, Beijing 100875, China
}

\date{Accepted XXX. Received YYY; in original form ZZZ}

\pubyear{2019}

\begin{document}
\label{firstpage}
\pagerange{\pageref{firstpage}--\pageref{lastpage}}
\maketitle

\begin{abstract}
Kilonova explosions typically release $\sim 10^{50-51}$ erg in kinetic energy, which is sufficient to constitute a kilonova remnant (KNR); however, it has not yet been confirmed. In this work, we investigate the probable association between G4.8+6.2 and the guest star of AD 1163, which is recorded by the Korea ancient astronomers. Although the evidence available is insufficient to draw a definite conclusion, it is at least theoretically self-consistent that the guest star of AD 1163 was a historical kilonova associated with G4.8+6.2, considering the possible short visible timescale of AD 1163, the relatively high Galactic latitude of G4.8+6.2, and that G4.8+6.2 is spatially coincident with the guest star of AD 1163. Further observation of G4.8+6.2 is needed to test our hypothesis. If our interpretation is correct, our results indicate that young KNRs should have a large diameter and low surface brightness, unlike other young supernova remnants.
\end{abstract}




\section{Introduction}

Kilonovae have been one of the most interesting topics in astrophysics since the detection of the gravitational wave event GW170817 and the accompanying kilonova AT 2017gfo \citep{Abbott2017}. However, the interpretation of a kilonova is still incomplete; for example, the ejecta properties are not understood \citep{Perego2017}. Several ejection mechanisms attempt to predict the observed behavior of a kilonova \citep{Cowperthwaite2017, Arcavi2018, Waxman2018, Kawaguchi2018}. For an unresolved point source, it is difficult to distinguish between these models directly from the light curve of a kilonova.

One way around this problem is to examine the remnant phase when a kilonova departs from free expansion and begins to strongly interact with the surrounding interstellar medium (ISM) \citep{Milisavljevic2017}. Kilonova remnants (KNRs) appear as extended sources, which in turn provide detailed three-dimensional kinematic and chemical information of the ejecta. Searching for young, nearby KNRs is necessary but difficult, not only because the observable lifetime of KNRs is shorter than that of supernova remnants (SNRs) \citep{Frail1994} but also because the event rate of kilonovae is much lower than that of supernovae \citep{Li1998}.

Galactic SNRs can be traced up to several tens of thousands of years ago \citep{Reynolds2012}. Estimates for the kilonova rate may up to $10^{-4}$ per year in the Milky Way due to the discovery of GW170817 during the total observation time of LIGO/VIRGO set O2 \citep{Lipunov2018}. Therefore, we expect one or more KNRs in the Milky Way, while none has yet been confirmed. This deficit motivated us to search for KNRs. Kilonovae from compact binary coalescence are generally expected to appear in the outskirts of galaxies or even outside galaxies where the density of the diffuse medium is low \citep{Avanzo2015}. This violent merger will eject some matter with a subrelativistic velocity. Kilonova explosions typically release $10^{50-51}$ erg in kinetic energy \citep{Li1998,Metzger2010}. We would expect a kilonova explosion to constitute an extended source similar to SNRs. In a low-density medium, the KNR exhibits both a low surface brightness and a rapid expansion to large sizes \citep{Tang2005}. Nonetheless, knowledge about KNRs is rather limited. Identifying the remnant as a potential KNR in all-sky surveys remains challenging.

Some early astronomical records may provide crucial clues for finding a potential KNR with the hope that we can extract the properties of a kilonova in early astronomical records. When the kilonova in the Milky Way galaxy was exploded, it might be taken as a guest star by the ancient astronomers. It may have different peak brightness, color, and duration of visibility, comparing with other phenomenon like a supernova. Motivated by this idea, we attempt to search KNRs in the sample of guest stars mainly drawn from \cite{Yau1988}.

In this work, we found a potential KNR candidate, G4.8+6.2, based on the description of AD 1163 and the properties of G4.8+6.2. The layout of this Letter is as follows. In Section. \ref{sec:Sample}, we briefly introduce the search method. In Section. \ref{sec:AD 1163}, we discuss the guest star of AD 1163. In Section. \ref{sec:SNR}, we discuss the probable kilonova remnant G4.8+6.2. The conclusions are summarized in Section. \ref{sec:summary}.

\section{Sample}\label{sec:Sample}

A kilonova is a transient astronomical event whose duration may last from days to weeks. The kilonova should also be considered as a prime candidate of the guest star, although a kilonova is rarer, dimmer and faster evolving than a normal supernova. It is difficult to establish a clear association between KNRs and the records, given the uncertainties of the derived KNR ages and the vague positions in ancient texts. Despite these difficulties, the guest star may provide crucial clues for finding potential young KNRs.

The major differences between the identification of novae, kilonovae, and supernovae are as follows. (1) The period of visibility to the naked eye for a typical supernova is usually longer than that for a nova or a kilonova. (2) The majority of SNRs are within one degree of the galactic equator, which is quite opposite to the distribution of novae or KNRs. (3) Kilonovae or supernovae are typically several kpc from the Earth, which is further than those of novae. (4) Novae can be seen at the same place multiple times, while kilonovae or supernovae cannot. (5) Unlike a nova, kilonova or supernova explosions may leave behind a remnant. (6) The Galactic event rate of novae is usually once every year, the event rate of supernovae is roughly once every fifty years, and kilonovae are detected roughly once every thousand years.

\begin{figure*}
    \includegraphics[width=\textwidth]{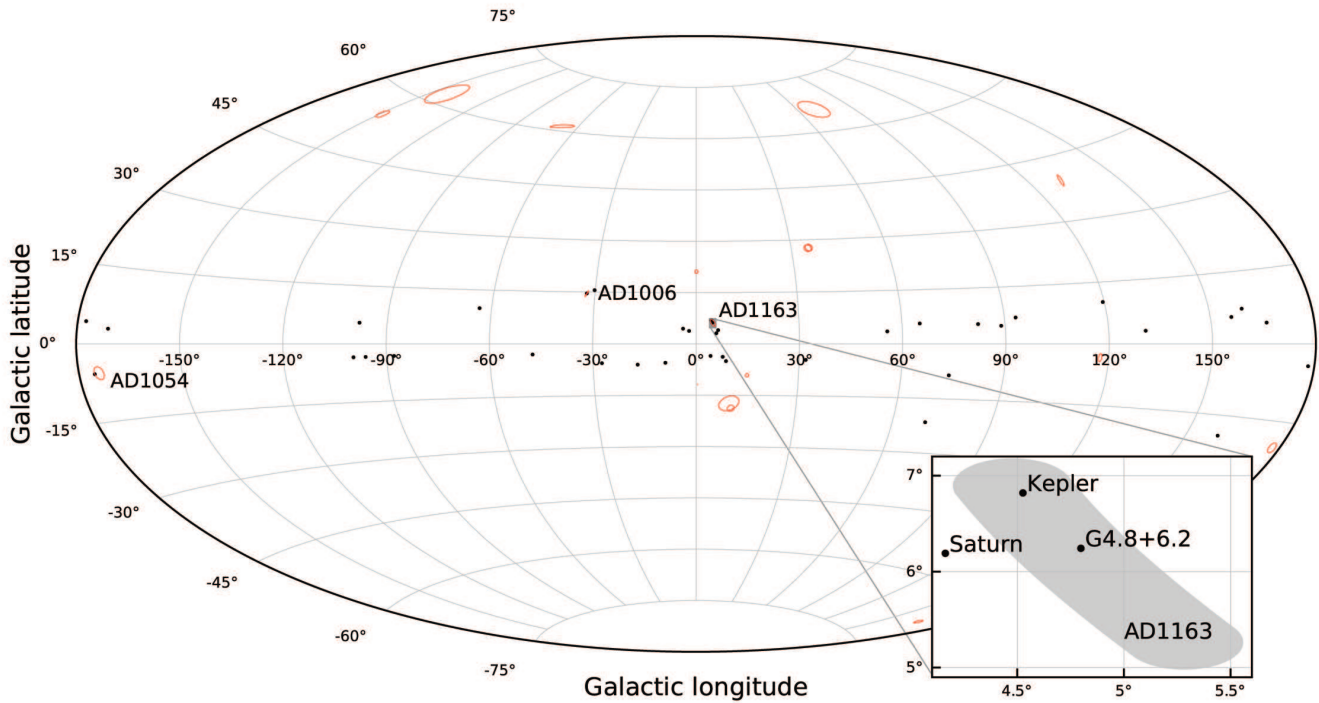}
    \caption{The sky distribution of 38 SNRs and 19 guest stars with latitudes larger than 3 degrees in the Galactic coordinate system. The black points mark the positions of the SNR in Green catalog \citep{Green2017}. The red circles correspond to the guest star with the error box listed in Table. \ref{tab:historic_record}. The inset plot shows the Moon's trajectory at the night of AD 1163 August 10 (in galactic coordinates), which is shown as the shadow region. Kepler and G4.8+6.2 was occulted by the Moon this evening, and Saturn was only about half degree from the Moon.}
    \label{fig:SNR}
\end{figure*}

Based on these differences, we first looked for a guest star with a short period of visibility and with high Galactic latitude. We list the historical records in the catalog of \cite{Yau1988} in Table. \ref{tab:historic_record}, where we excluded the sources with Galactic latitudes less than three degrees. To obviate an accidental agreement in position, we do not consider the guest stars with error bars larger than three degrees. We attempt to identify the potential counterparts of kilonova candidates by the association between the guest star and its remnant. We show the result in Figure. \ref{fig:SNR}. Only three have remnants located inside the error boxes of the guest stars. Two are the well-known AD 1006 and AD 1054 associated with SNR G327.6+14.6 and SNR G184.6-5.8, respectively. The third is the guest star of AD 1163 located near G4.8+6.2, which is the most preferable candidate of a kilonova.

\section{The guest star of AD 1163}\label{sec:AD 1163}

The guest star of AD 1163 observed in Korea was recorded in a rather unique way. The record says: ``On AD 1163 August 10 a guest star invaded the Moon."\citep{Stephenson1971}. Here we calculate the Moon's trajectory at the night of AD 1163 August 10 with a public code PyEphem \footnote{https://rhodesmill.org/pyephem/index.html} (in galactic coordinates), which should be the coordinates of the guest star of AD 1163 (see the inset plot of Figure. \ref{fig:SNR}). With the hypothesis of the conjunction between the guest star and the Moon, \cite{Stephenson1971} suggests that AD 1163 must have been visible for several days and that the apparent magnitude can be estimated to be brighter than 1. One can see more details in \cite{Stephenson1971}. However, the exact duration cannot be inferred from the record above.

The guest star of AD 1163 was recorded in the Korean historical book, while it was not shown in Chinese historical book. However, ``Songshi Tianwenzhi", which is an officially edited history book for Song Dynasty, has a record about the Moon invading Saturn around the same time. We have translated the Chinese records as follows:

``Longxing reign period, 1st year, 3th month, day bingshen [33]; 4th month, day bingzi [13]; 7th month, day wuxu [35]; Moon trespassed against Saturn." [Songshi, vol. 53, Tianwenzhi No. 6]. Dates, converted to the Julian calendar from the tables of \cite{Fang2007}, are respectively 1163 April 10, May 20, and August 10.

We also calculated the position of the Saturn on the evening of 1163 August 10 (in galactic coordinates) with PyEphem. The results are shown in the inset plot of Figure. \ref{fig:SNR}, which is consistent with the results presented in \cite{Stephenson1971}. Stephenson has discussed in detail and ruled out the possibility that Saturn was mistaken for a guest star by the Korean astronomers \citep{Stephenson1971, Stephenson2009}. But the reason why the Chinese astronomers missed the observation of AD 1163 is still unclear.

\begin{figure}
\includegraphics[width=\columnwidth]{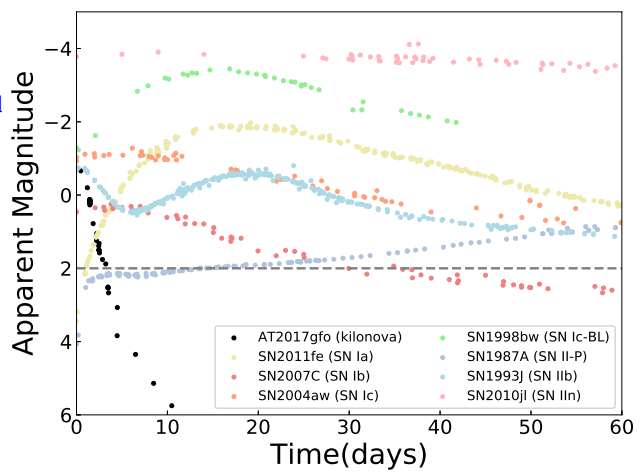}
\caption{V-band light curves of GW170817 at 10 kpc in comparison to different types of typical supernovae at 15 kpc. The black horizontal line indicates an apparent magnitude equals to 2 mag. }
\label{fig:optical}
\end{figure}

In the following, we show the time scale of visibility of a kilonova by taking AT 2017gfo as a typical kilonova. AT 2017gfo with an absolute peak magnitude of $M = -15.8 \pm 0.1$ and a r-band decline rate of 1.1 mag/d offers the unique opportunity to study kilonova emission that was plausibly not polluted by the gamma ray burst (GRB) afterglow emission \citep{Valenti2017, Covino2017}.  We put AT 2017gfo at a distance of 10 kpc and compare it with some typical supernovae at 15 kpc (see detailed reason why 10 kpc and 15 kpc are adopted in Section. \ref{sec:age}). In Figure. \ref{fig:optical} we show the V-band light curve chosen to investigate visible timescale of KNRs and SNRs. All photometric data of supernovae was obtained from The Open Supernova Catalog \citep{Guillochon2017} \footnote{\url{https://sne.space}}.
The peak apparent magnitude of AT 2017gfo will be approximately $-1~\rm mag$ and the observable time scale for naked eyes is around 3 days, where we choose the naked-eye limiting magnitudes to be $2~\rm mag$. It is obvious that the observable time scale for typical supernovae at 15 kpc would be much longer. This short duration suggests that AD 1163 is more likely to be a nova or kilonova rather than a typical supernova, although a fast-evolving type of supernova \citep{Poznanski2010, Chen2019} cannot be completely ruled out based on current evidence. Unlike novae, kilonovae may leave behind a remnant after an explosion.

AD 1163 was initially regarded as a nova. \cite{Stephenson1971} argued that the guest star of AD 1163 may have been a previous outburst of the type-Ia SN Kepler of AD 1604 \citep{Baade1943, Neuhauser2016, Reynolds2007}. \cite{Yau1988} argued that AD 1163 could have been a periodic outburst with a period of approximately 800 years, while no other observations being reported. Notice that G4.8+6.2 ($RA_{J2000} = 17h33m24s$, $Dec_{J2000} = -21d34m$) lies 40 arcmin east from Kepler's SNR and appears within the region of sky obscured by the Moon on the night of AD 1163 August 10 \citep{Stephenson1971, Bhatnagar2000}, as shown in the inset plot of Figure. \ref{fig:SNR}. We suggest that G4.8+6.2 is a KNR associated with the guest star AD 1163.

\section{Supernova Remnant G4.8+6.2}\label{sec:SNR}

\subsection{Age estimate}\label{sec:age}

G4.8+6.2 is a typical shell-type SNR with a spectral index of $\alpha = -0.57 \pm 0.13$ and a nearly circular structure of size $17 \times 18$ $arcmin^2$ \citep{Bhatnagar2000}. The distance to G4.8+6.2 is estimated to be approximately 15 kpc based on the $\Sigma - D$ relationship \citep{Bhatnagar2000}. The error in this estimate is larger than 75\% and dominated by the $\Sigma - D$ relation itself \citep{Bhatnagar2000}. Furthermore, the distance estimate is dependent on the assumption that it is an SNR. As shown in eq. (10) of \cite{Duric1986}, the surface brightness depends positively on the kinetic energy and the density of the environment, which are both thought smaller for KNRs. KNRs would tend to have, on average, lower surface brightness in comparison to SNRs. The $\Sigma - D$ relation for two distinctive classes of remnants, SNRs and KNRs, have two tracks or domains in the $\Sigma - D$ plane, one above the other \citep{Arbutina2005}. With same surface brightness $\Sigma$, the diameter $D$ for KNR might be smaller.

The shock radius $r_{s}$ is $39 \times (\frac{d}{15 \rm kpc})$ pc corresponding to an angular diameter of $\sim 18$ arcmin, where d is the distance to G4.8+6.2. We infer the age of G4.8+6.2 by applying the Sedov model \citep{Jiang2007, Sedov1959}, which is well described by the self-similar solution for a point explosion in a uniform medium. Considering the possibility of the association with AD 1163 and the pre-explosion environment of a KNR, we assume an initial explosion energy of $10^{51} {\rm ergs}$ and an average ambient density of $n_{0} \sim 0.001 {\rm cm^{-3}}$. These typical values are based on the following considerations. The presented simulations show that the kinetic energy of a kilonova is comparable to that of a supernova \citep{Rosswog2013, Wu2019}. The maximum expected fractional polarization for a uniform magnetic field is approximately 70.2\%, by assuming the free electron energy spectral index $\gamma = 2.14$ \citep{Reynolds1993}, while the percentage polarization of G4.8+6.2 increases to 70\% at 1400 MHz (NVSS) near the edge of the remnant \citep{Duncan1997, Zhang2003}. The unusual polarization properties indicate that G4.8+6.2 may be a nearby remnant that was born in an environment of a very-low-density ISM due to the effect of depolarization \citep{Zhang2003}. But it is worth noticing that G4.8+6.2 have high latitudes, so that small depolarization may be due to G4.8+6.2 located high above the Galactic plane.

We obtain a dynamical age of $t=2544 \times\left(\frac{n_{0}}{0.001 \mathrm{cm}^{-3}}\right)^{\frac{1}{2}} \times \left(\frac{r_{s}}{39 {\rm pc}}\right)^{\frac{5}{2}} \times\left(\frac{E}{10^{51} {\rm erg}}\right)^{-\frac{1}{2}} {\rm yr}$ by adopting the Sedov solution \citep{Sedov1959}. We mentioned that G4.8+6.2 may not satisfy the $\Sigma - D$ relation if it was indeed a KNR, and hence, the distance was overestimated. G4.8+6.2 is in the direction toward the galactic center. Therefore, we believe G4.8+6.2 is not too far away from the Earth. Otherwise, it will be in the opposite direction of the galactic center and the depolarization effect must be important. A more reliable measurement of the distance is important. Considering the uncertainties of the distance, a 900-year-old KNR could be shown as G4.8+6.2, e.g., with a distance of 10 kpc, and the dynamical age could be as low as 900 years, which is consistent with the guest star observed on AD 1163 August 10, and is also consistent with KNR's $\Sigma-D$ relation. Therefore, we believe that the association between G4.8+6.2 and AD 1163 is quite promising.

We also attempt to use the polarization information to estimate the age of G4.8+6.2. The magnetic field is expected to be radial in young shell-type SNRs and tangential in old SNRs \citep{Reynolds2012, Woltjer1972}. However, the map of the polarization position angles at 1400 MHz shows that the orientation of the magnetic field does not align in the radial direction or the tangential direction \citep{Zhang2003}.

\subsection{Image}\label{sec:image}

The morphological information from hydrodynamic simulations may offer clues to explain the difference between KNRs and SNRs with different initial and boundary conditions because the morphologies are shaped in large part by their explosions and environments \citep{Lopez2018}. The ejecta of a kilonova from the progenitor are expected to be much more anisotropic than that of a supernova. In addition, the progenitor of a KNR generally resides within approximately uniform surrounding material.

\citet{Zhang2003} mentioned that G4.8+6.2 appears as a barrel (the radio emission is obviously weak on the north and south edges), as shown in Fig. 1 of \citet{Zhang2003}.
The lack of alignment between the symmetric axes of the bilateral SNRs and the galactic plane rules out the use of the magnetic models proposed by \citet{Shaver1969} and \citet{Gaensler1998} to explain the barrel shape. It is possible that the shape of G4.8+6.2 could be explained by ``intrinsic" explosion models, because young KNRs are not highly contaminated by the swept-up interstellar medium and therefore provide plenty of information about kilonovae. The tidal ejecta mainly concentrate in the equatorial plane as expected for the merging of a double neutron star \citep{Kasen2017}, while the ambient density is likely uniform, as indicated by the spherical radio morphology.

\begin{figure}
	\includegraphics[width=\columnwidth]{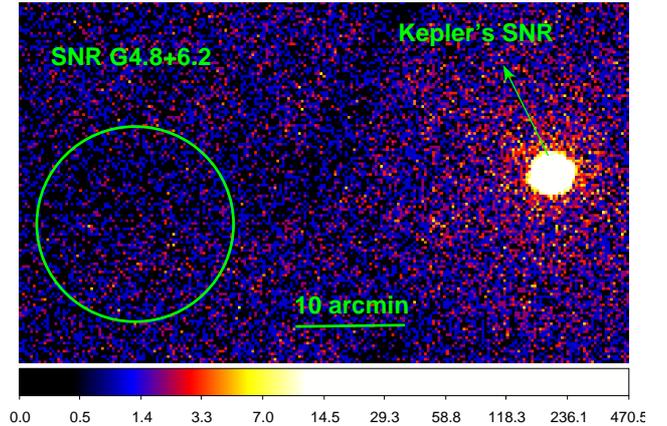}
    \caption{X-ray counts image of G4.8+6.2. The data are obtained from the ROSAT instrument. The region of G4.8+6.2 is indicated by a green circle with a radius of 9 arcmin. The color bar shows the logarithm scale of the image.  G4.8+6.2 does not have prominent X-ray emission and is strongly contaminated by Kepler's SNR which is located toward G4.8+6.2's west.}
    \label{fig:X-ray}
\end{figure}

As indicated in Figure. \ref{fig:X-ray}, the X-ray emission from G4.8+6.2 is weak and is contaminated heavily by Kepler's SNR. However, many young shell SNRs show evidence for X-ray emission. This might be because the total energy of a kilonova is smaller than that of a supernova and surrounding medium is tenuous or because KNRs evolve faster than SNRs. If G4.8+6.2 was in a very-low-density environment, it would have resulted in extremely low X-ray emission. A low-density medium appears to have no effect on the radio emission \citep{Landecker1999}. No optical emission was detected towards G4.8+6.2, may be due to the obscuration of the central regions of the galactic plane. The unclear nature of both the optical and X-ray observations does not help to estimate the distance of G4.8+6.2.

\section{Summary and Future Prospects}\label{sec:summary}
In the letter, we attempted to search for a possible KNR in our own galaxy. We used guest star information to reduce the search sample and found a potential KNR candidate G4.8+6.2. We discussed the probable association between G4.8+6.2 and AD 1163 from different aspects. The position of G4.8+6.2 is consistent with that of the guest star of AD 1163, and the size properties suggest that its age is also comparable. The possible duration of AD 1163 and the environment of G4.8+6.2 are consistent with those of the model of a kilonova. If our interpretation is correct, our results indicate that young KNRs should have a large diameter and low surface brightness, unlike other young SNRs. A further investigation of G4.8+6.2 and its possible KNR would be of great interest, since young KNRs are less contaminated by the swept-up interstellar medium and therefore provide more information about the kilonova.

However we need to be cautious that the association among AD 1163, G4.8+6.2 and a kilonova interpretation are still speculative. It remains unclear why AD1163 was only recorded by Korean astronomers but not Chinese astronomers, despite the fact that Chinese astronomers recorded the moon invading Saturn on the same date. What is more, the duration of the AD1163 was not directly recorded in any ancient records. Even the AD 1163 - G4.8+6.2 association is true, it may also be due to a fast-evolving type of supernova rather than a kilonova. Given the currently available information, it is difficult to firmly determine the nature of G4.8+6.2/AD1163. An accurate measurement of G4.8+6.2 could help clarify this issue. If G4.8+6.2 is indeed a young kilonova remnant as we suggested, the forward shock speed is expected to be higher, which could be tested with future observations by comparing radio images at epochs separated by more than a decade. The radioactive decay of freshly synthesized r-process nuclei can also help us to examine whether G4.8+6.2 is a true kilonova remnant in the future \citep{Wu2019, Korobkin2019}.

We also notice that the guest stars of AD 1356 and AD 1399 were recorded in a similar way as that of AD 1163 \citep{Stephenson1971}. However, the galactic latitude of AD 1356 is less than 3 degrees, and no remnants were observed to be close to the position of AD 1399.

The identification of Galacitc KNR gives promising constraints on the population of close binaries. With relatively short orbital period, they could be ideal gravitational wave candidates for the space-borne detectors, such as Tianqin \citep{Luo2016} and LISA \citep{2017arXiv170200786A}. These candidates are ideal objects with multi-messenger (both in electromagnetic wave and in gravitational wave) and multi-wavelength observations.

\begin{table*}
	\centering
	\caption{historic records of new stars with galactic latitudes larger than 3 degrees. The data are taken from \citep{Yau1988}.}
	\label{tab:historic_record}
	\begin{tabular}{llllllll} 
		\hline
		Date & RA & RA\_error & Dec & Dec\_error & Duration & Other name \\
		\hline
		B.C. 104-101    &    14h30m &      12m &    +39d &        3d &      ... &           ...\\
		B.C. 77 Oct/Nov &     11h4m &       6m &     76d &      1.5d &      ... &           ...\\
		B.C. 48 Hay/Jun &    19h23m &       4m &    -28d &        1d &      ... &    G21.5-0.9?\\
		A.D. 29         &    17h15m &       4m &    +14d &        1d &      ... &           ...\\
		101 Dec 30      &    10h17m &       4m &    +23.4d &        1d &      ... &  PG 0917+342?\\
		222 Nov 4       &    12h31m &      11m &   -1.1d &      0.4d &      ... &           ...\\
		304 Jun/Jul     &     4h29m &       4m &    +19d &        1d &      ... &           ...\\
		642 Aug 9       &    16h36m &       ... &    -19d &         ...  &      ... &           ...\\
		829 Nov/Dec     &    10h52m &      12m &    +22d &        3d &      ... &           ...\\
		891 May 11      &    16h31m &       2m &  -16.6d &      0.5d &      ... &           ...\\
		911 May/Jun     &    17h15m &       4m &   14.4d &        1d &      ... &           ...\\
		1006 Apr 3      &     15h3m &     0.5m &    -42d &        1d &    16 mn &   G327.6+14.6\\
		1011 Feb 8      &    19h16m &       9m &    -28d &        3d &      ... &           ...\\
		1054 May 20     &     5h38m &       5m &    +21d &      1.5d &     2 yr &    G184.6-5.8\\
		1163 Aug 10     &    17h34m &       5m &  -21.5d &      0.5d &      ... &     V794 Oph?\\
		1399 Jan 5      &    18h52m &       2m &  -20.1d &      0.5d &      ... &           ...\\
		1592 Nov 28     &     1h51m &       2m &  -10.4d &      0.5d &      ... &           ...\\
		1592 Dec 4      &      0h9m &       4m &  +59.1d &        1d &    15 mn &           ...\\
		1690 Sep 29     &  18h35.5m &     0.2m &  -34.1d &      0.1d &     2 dy &           ...\\
		\hline
	\end{tabular}
\end{table*}

\section*{Acknowledgements}

We thank the anonymous referee for the helpful comments which have helped us to improve the presentation
of the paper. Y.Z. was supported in part by Perimeter Institute for Theoretical Physics. Research at Perimeter Institute is supported by the Government of Canada through the Department of Innovation, Science and Economic Development Canada and by the Province of Ontario through the Ministry of Economic Development, Job Creation and Trade. This work was supported by the National Natural Science Foundation of China (Grant No. U1738132, 11690024, 11203013 and 11722324). B.J. acknowledges the Jiangsu NSF grant BK20141310 and SRFDP of China 20110091120001.





\begin{thebibliography}{99}
\bibitem[\protect\citeauthoryear{Abbott, et al.}{2017}]{Abbott2017} Abbott B.~P., et al., 2017, ApJL, 848, L12
\bibitem[\protect\citeauthoryear{Amaro-Seoane, et al.}{2017}]{2017arXiv170200786A} Amaro-Seoane P., et al., 2017, arXiv e-prints, arXiv:1702.00786
\bibitem[\protect\citeauthoryear{Arbutina \& Uro{\v{s}}evi{\'c}}{2005}]{Arbutina2005} Arbutina B., Uro{\v{s}}evi{\'c} D., 2005, MNRAS, 360, 76
\bibitem[\protect\citeauthoryear{Arcavi}{2018}]{Arcavi2018} Arcavi I., 2018, ApJL, 855, L23
\bibitem[\protect\citeauthoryear{Baade}{1943}]{Baade1943} Baade W., 1943, ApJ, 97, 119
\bibitem[\protect\citeauthoryear{Bhatnagar}{2000}]{Bhatnagar2000} Bhatnagar S., 2000, MNRAS, 317, 453
\bibitem[\protect\citeauthoryear{Chen, et al.}{2019}]{Chen2019} Chen P., et al., 2019, arXiv, arXiv:1905.02205
\bibitem[\protect\citeauthoryear{Covino, et al.}{2017}]{Covino2017} Covino S., et al., 2017, Nature Astronomy, 1, 791
\bibitem[\protect\citeauthoryear{Cowperthwaite, et al.}{2017}]{Cowperthwaite2017} Cowperthwaite P.~S., et al., 2017, ApJL, 848, L17
\bibitem[\protect\citeauthoryear{D'Avanzo}{2015}]{Avanzo2015} D'Avanzo P., 2015, Journal of High Energy Astrophysics, 7, 73
\bibitem[\protect\citeauthoryear{Duncan, Stewart, Haynes \& Jones}{1997}]{Duncan1997} Duncan A.~R., Stewart R.~T., Haynes R.~F., Jones K.~L., 1997, MNRAS, 287, 722
\bibitem[\protect\citeauthoryear{Duric \& Seaquist}{1986}]{Duric1986} Duric N., Seaquist E.~R., 1986, ApJ, 301, 308
\bibitem[\protect\citeauthoryear{Fang, et al.}{2007}]{Fang2007}
Fang S., Fang X., 2007, Zhongguo shi li ri he Zhong xi li ri dui zhao biao, di 1 ban edn. Shanghai ren min chu ban she, Shanghai
\bibitem[\protect\citeauthoryear{Frail, Goss \& Whiteoak}{1994}]{Frail1994} Frail D.~A., Goss W.~M., Whiteoak J.~B.~Z., 1994, ApJ, 437, 781
\bibitem[\protect\citeauthoryear{Gaensler}{1998}]{Gaensler1998} Gaensler B.~M., 1998, ApJ, 493, 781
\bibitem[\protect\citeauthoryear{Green}{2017}]{Green2017} Green D.~A., 2017, yCat, VII/278
\bibitem[\protect\citeauthoryear{Guillochon, Parrent, Kelley \& Margutti}{2017}]{Guillochon2017} Guillochon J., Parrent J., Kelley L.~Z., Margutti R., 2017, ApJ, 835, 64
\bibitem[\protect\citeauthoryear{Jiang, Chen \& Wang}{2007}]{Jiang2007} Jiang B., Chen Y., Wang Q.~D., 2007, ApJ, 670, 1142
\bibitem[\protect\citeauthoryear{Kasen, Metzger, Barnes, Quataert \& Ramirez-Ruiz}{2017}]{Kasen2017} Kasen D., Metzger B., Barnes J., Quataert E., Ramirez-Ruiz E., 2017, Nature, 551, 80
\bibitem[\protect\citeauthoryear{Kawaguchi, Shibata \& Tanaka}{2018}]{Kawaguchi2018} Kawaguchi K., Shibata M., Tanaka M., 2018, ApJL, 865, L21
\bibitem[\protect\citeauthoryear{Korobkin, et al.}{2019}]{Korobkin2019} Korobkin O., et al., 2019, arXiv e-prints, arXiv:1905.05089
\bibitem[\protect\citeauthoryear{Landecker, Routledge, Reynolds, Smegal, Borkowski \& Seward}{1999}]{Landecker1999} Landecker T.~L., Routledge D., Reynolds S.~P., Smegal R.~J., Borkowski K.~J., Seward F.~D., 1999, ApJ, 527, 866
\bibitem[\protect\citeauthoryear{Li \& Paczy{\'n}ski}{1998}]{Li1998} Li L.-X., Paczy{\'n}ski B., 1998, ApJ, 507, L59
\bibitem[\protect\citeauthoryear{Lipunov, et al.}{2018}]{Lipunov2018} Lipunov V., et al., 2018, NewA, 63, 48
\bibitem[\protect\citeauthoryear{Lopez \& Fesen}{2018}]{Lopez2018} Lopez L.~A., Fesen R.~A., 2018, SSRv, 214, 44
\bibitem[\protect\citeauthoryear{Luo, et al.}{2016}]{Luo2016} Luo J., et al., 2016, CQGra, 33, 35010
\bibitem[\protect\citeauthoryear{Milisavljevic \& Fesen}{2017}]{Milisavljevic2017} Milisavljevic D., Fesen R.~A., 2017, Handbook of Supernovae,  2211
\bibitem[Metzger et al.(2010)]{Metzger2010} Metzger, B.~D., Mart{\'\i}nez-Pinedo, G., Darbha, S., et al.\ 2010, MNRAS, 406, 2650
\bibitem[\protect\citeauthoryear{Neuh{\"a}user, Rada, Kunitzsch \& Neuh{\"a}user}{2016}]{Neuhauser2016} Neuh{\"a}user R., Rada W., Kunitzsch P., Neuh{\"a}user D.~L., 2016, JHA, 47, 359
\bibitem[\protect\citeauthoryear{Perego, Radice \& Bernuzzi}{2017}]{Perego2017} Perego A., Radice D., Bernuzzi S., 2017, ApJL, 850, L37
\bibitem[\protect\citeauthoryear{Poznanski, et al.}{2010}]{Poznanski2010} Poznanski D., et al., 2010, Science, 327, 58
\bibitem[\protect\citeauthoryear{Reynolds \& Gilmore}{1993}]{Reynolds1993} Reynolds S.~P., Gilmore D.~M., 1993, AJ, 106, 272
\bibitem[\protect\citeauthoryear{Reynolds, Borkowski, Hwang, Hughes, Badenes, Laming \& Blondin}{2007}]{Reynolds2007} Reynolds S.~P., Borkowski K.~J., Hwang U., Hughes J.~P., Badenes C., Laming J.~M., Blondin J.~M., 2007, ApJ, 668, L135
\bibitem[\protect\citeauthoryear{Reynolds, Gaensler \& Bocchino}{2012}]{Reynolds2012}Reynolds S.~P., Gaensler B.~M., Bocchino F., 2012, SSRv, 166, 231
\bibitem[\protect\citeauthoryear{Rosswog}{2013}]{Rosswog2013} Rosswog S., 2013, Philosophical Transactions of the Royal Society of London Series A, 371, 20120272
\bibitem[\protect\citeauthoryear{Sedov}{1959}]{Sedov1959} Sedov L.~I., 1959, New York: Academic Press
\bibitem[\protect\citeauthoryear{Shaver}{1969}]{Shaver1969} Shaver P.~A., 1969, The Observatory, 89, 227
\bibitem[\protect\citeauthoryear{Shen}{1088}]{Shen1088} Shen, K.\ 1088, Meng Xi Bi Tan, Xiang Shu 2 (English translation: Brush Talks from Dream Brook (two volumes), by Wang Hong and Zhao Zheng, published in 2008 by Sichuan People's Publishing House, China.)
\bibitem[\protect\citeauthoryear{Stephenson}{1971}]{Stephenson1971} Stephenson F.~R., 1971, ApL, 9, 81
\bibitem[\protect\citeauthoryear{Stephenson \& Green}{2009}]{Stephenson2009} Stephenson F.~R., Green D.~A., 2009, JHA, 40, 31
\bibitem[\protect\citeauthoryear{Tang \& Wang}{2005}]{Tang2005} Tang S., Wang Q.~D., 2005, ApJ, 628, 205
\bibitem[\protect\citeauthoryear{Valenti, et al.}{2017}]{Valenti2017} Valenti S., et al., 2017, ApJL, 848, L24
\bibitem[\protect\citeauthoryear{Waxman, Ofek, Kushnir \& Gal-Yam}{2018}]{Waxman2018} Waxman E., Ofek E.~O., Kushnir D., Gal-Yam A., 2018, MNRAS, 481, 3423
\bibitem[\protect\citeauthoryear{Woltjer}{1972}]{Woltjer1972} Woltjer L., 1972, ARA\&A, 10, 129
\bibitem[\protect\citeauthoryear{Wu, et al.}{2019}]{Wu2019} Wu M.-R., et al., 2019, arXiv e-prints, arXiv:1905.03793
\bibitem[\protect\citeauthoryear{Yau}{1988}]{Yau1988} Yau K.~K.~C., 1988, PhD Thesis, Durham Univ. 
\bibitem[\protect\citeauthoryear{Zhang, Strom \& Reich}{2003}]{Zhang2003} Zhang X.-Z., Strom R.~G., Reich W., 2003, Chinese Physics Letters, 20, 969
\end{thebibliography}


\bsp	
\label{lastpage}
\end{document}